# Fourier Optics in the Classroom

Masud Mansuripur, College of Optical Sciences, the University of Arizona, Tucson

**Abstract**: Borrowing methods and formulas from Prof. Goodman's classic *Introduction to Fourier Optics* textbook[1], I have developed a software package[2] that has been used in both industrial research and classroom teaching.[3] This paper briefly describes a few optical system simulations that have been used over the past 30 years to convey the power and the beauty of Fourier Optics to our students at the University of Arizona's College of Optical Sciences.

**Introduction**. The following examples are representative computer simulations that show how Fourier optics, when combined with traditional methods of geometric optics (i.e., ray-tracing) and physical optics (e.g., exact solutions of Maxwell's equations), can be used in the classroom to demonstrate not only the physical principles but also real world applications of optical science and engineering. Details of these simulations as well as numerous other examples can be found in Ref.[3].

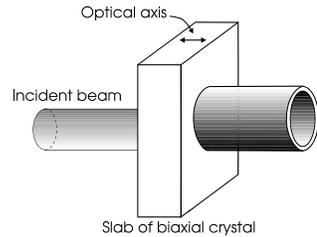

**Example 1: Internal conical refraction**. The figure on the right shows a collimated beam ($\lambda = 0.65\mu m$) entering a slab of biaxially birefringent crystal. The crystal is cut with one of its optical axes perpendicular to the surface. Upon entering the slab, the beam spreads into a cone, and emerges in the form of two bright, concentric cylinders, as shown in Figs.1 and 2 below.

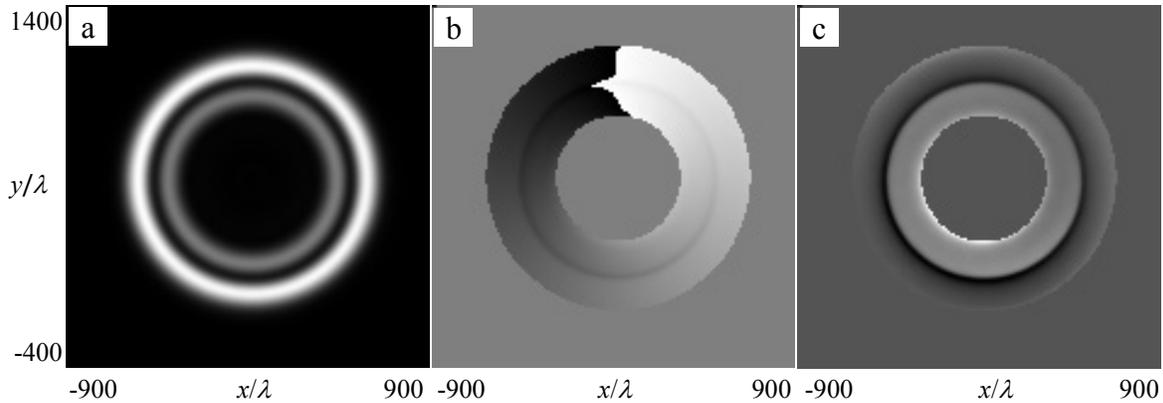

**Fig.1**. Distributions at the exit facet of the crystal slab for a circularly polarized incident beam. (a) Total intensity $I_x + I_y$; (b) polarization rotation angle $\rho$; (c) polarization ellipticity $\eta$.

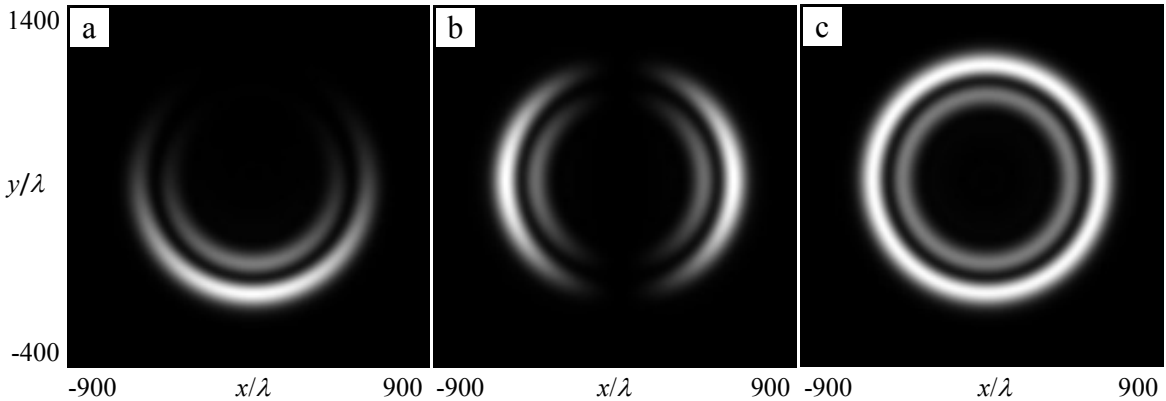

**Fig.2**. Intensity distributions for (a) $x$-component and (b) $y$-component of polarization at the exit facet of the crystal when the incident beam is linearly polarized along $x$-axis. (c) Distribution of total intensity $I_x + I_y$ at the exit facet of the crystal slab for an unpolarized incident beam.

**Example 2: Focusing polarized light through a slab of aragonite crystal**. Figure 3 illustrates the interaction between a focused laser beam ($\lambda = 633$ nm) and a biaxially birefringent crystal of aragonite. The microscope consists of a pair of linear polarizers, a beam-splitter, and an objective lens (NA = 0.375, $f =$ 12.66 mm) that focuses the beam in the forward path, and collimates the reflected light in the return path.

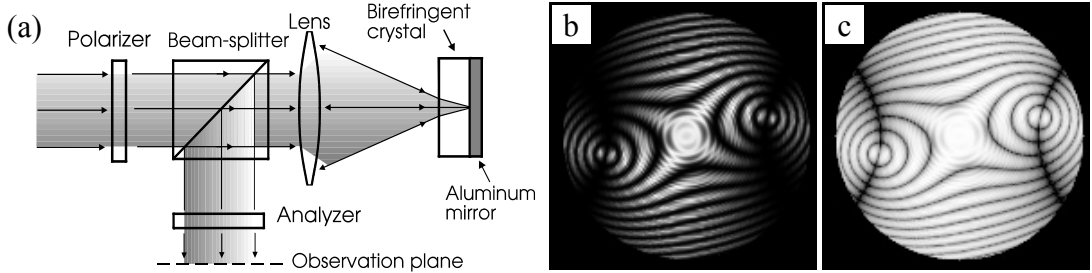

**Fig.3**. (a) Diagram of the simulated system. Double passage of the focused laser beam through the birefringent crystal slab causes variation of the polarization rotation angle over the beam's cross-section. Crossed polarizers convert polarization rotation into intensity variation. (b,c) Distributions of intensity and *log*(intensity) at the observation plane.

**Example 3: Diffractive optical elements coated on an axicon**. Figure 4 shows an axicon, coated with transparent DOEs on both its entrance and exit facets. The system parameters are given in the caption, and the intensity and phase profiles of the emergent beam at the destination plane are shown in Fig.5.

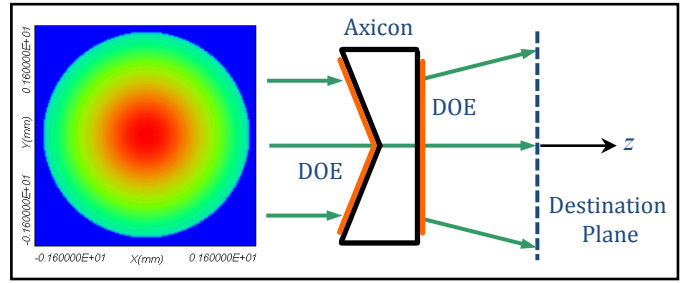

**Fig.4**. A Gaussian beam (diameter = 3.0 mm, $e^{-1}$ radius = 2.0 mm, $\lambda_0 = 0.65$ μm) passes through an axicon having a full cone angle of 179° and an aperture diameter of 4.0 mm. The DOE coating on the axicon's entrance facet has a phase profile $F(r) = r^2 - 0.6r^3$ (where $r$ is in mms). On the exit facet, the DOE phase profile is given by $F(x\,y) = -½(x^2 + 3y^2)$, where $x$ and $y$ are in units of mm, and where the $xy$-axes are rotated 45° relative to the $xy$ coordinate system. The construction wavelength of both DOEs is $\lambda_c = 0.65$μm, and the traced beam is the $+1^{st}$ diffracted order. The incident beam is linearly polarized along the $x$-axis, the refractive index and thickness of the axicon are 1.78 and 2.0 mm, respectively, and the destination plane is positioned 3.0mm past the axicon's flat surface.

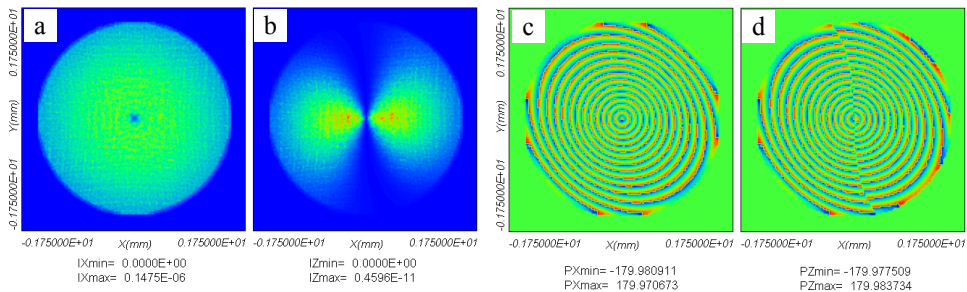

**Fig.5**. Plots of intensity (a,b) and phase (c,d) at the destination plane of the system depicted in Fig.4. (a and c) $x$-component of polarization; (b and d) $z$-component of polarization. The $z$-component is nearly five orders-of-magnitude weaker than the $x$-component.